\documentclass{article}
\usepackage{spconf,amsmath,graphicx}
\usepackage{amsmath}
\usepackage{mathtools}
\usepackage{amsfonts}
\usepackage{array}
\usepackage{cite}
\usepackage{amsthm}
\usepackage{balance}
\usepackage{color}
\usepackage{booktabs}
\usepackage{multirow}
\usepackage[font=small,labelfont=bf]{caption}

\newcommand{\argmin}{\mathrm{argmin}}
\title{Morphological Reconstruction of Detached Dendritic Spines via  Geodesic Path Prediction}
\name{Sammit Jain$^{1,\dagger}$, Suvadip Mukherjee$^{1,\dagger}$, Lydia Danglot$^2$ and Jean-Christophe Olivo-Marin$^{1,*}$
\thanks{$^\dagger$ Equally contributing authors. $^*$Corresponding author. S. Jain was a visiting scholar at Institut Pasteur, Paris. This work was supported by the Paris Ile-de-France Region through the DIM ELICIT program no. 1842.
\newline This work has been submitted to the IEEE for possible publication. Copyright may be transferred without notice, after which this version may no longer be accessible.}
}
\address{$^1$Institut Pasteur, BioImage Analysis Unit, Paris, France \\ $^2$Inserm U894, Center for Psychiatry \& Neuroscience, Paris, France}

\begin{document}
%
\maketitle
\begin{abstract}
Morphological reconstruction of dendritic spines from fluorescent microscopy is a critical open problem in neuro-image analysis. Existing segmentation tools are ill-equipped to handle thin spines with long, poorly illuminated neck membranes. We address this issue, and introduce an unsupervised path prediction technique based on a stochastic framework  which seeks the optimal solution from a \textit{path-space} of possible spine neck reconstructions. Our method is specifically designed to reduce bias due to outliers, and is adept at reconstructing challenging shapes from images plagued by noise and poor contrast.
Experimental analyses on two photon microscopy data demonstrate  the efficacy of our method,
where an improvement of $12.5\%$ is observed over the state-of-the-art in terms of mean absolute reconstruction error.
\end{abstract}
\begin{keywords}
dendritic spines, segmentation, active contour, geodesic shortest path, microscopy
\end{keywords}
%
\section{INTRODUCTION}
\label{sec:intro}
\begin{figure}[t]
	\setlength{\tabcolsep}{2pt}
	\centering
	\begin{tabular}{ccc}
		\includegraphics[width=.32\linewidth]{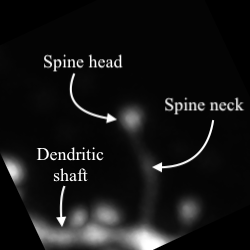} &
		\includegraphics[width=.322\linewidth]{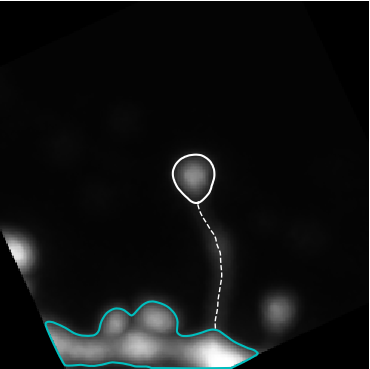} &
		\includegraphics[width=.322\linewidth]{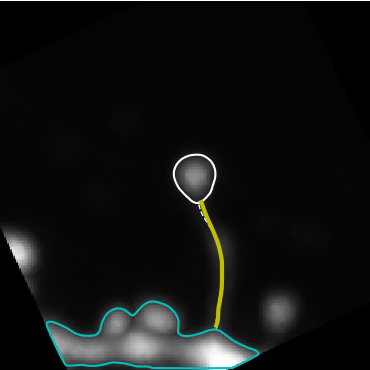} \\
		\small{(a)} & \small{(b)} & \small{(c)} 
	\end{tabular}
	\caption[illustration]{(a) An example two photon sub-image of a dendritic shaft and a \textit{thin} spine (b) Segmentation of the spine head and dendritic shaft via L2S\cite{l2s} are shown via white and cyan contours respectively. The manually annotated spine-neck path is shown in white dotted curve (c) Predicted spine-neck shown by the yellow curve.} 
	\label{fig:fig1}
	\vspace{-16pt}
\end{figure}

\begin{figure}[t]
	\centering
	\begin{tabular}{c}
		\includegraphics[width=.95\linewidth]{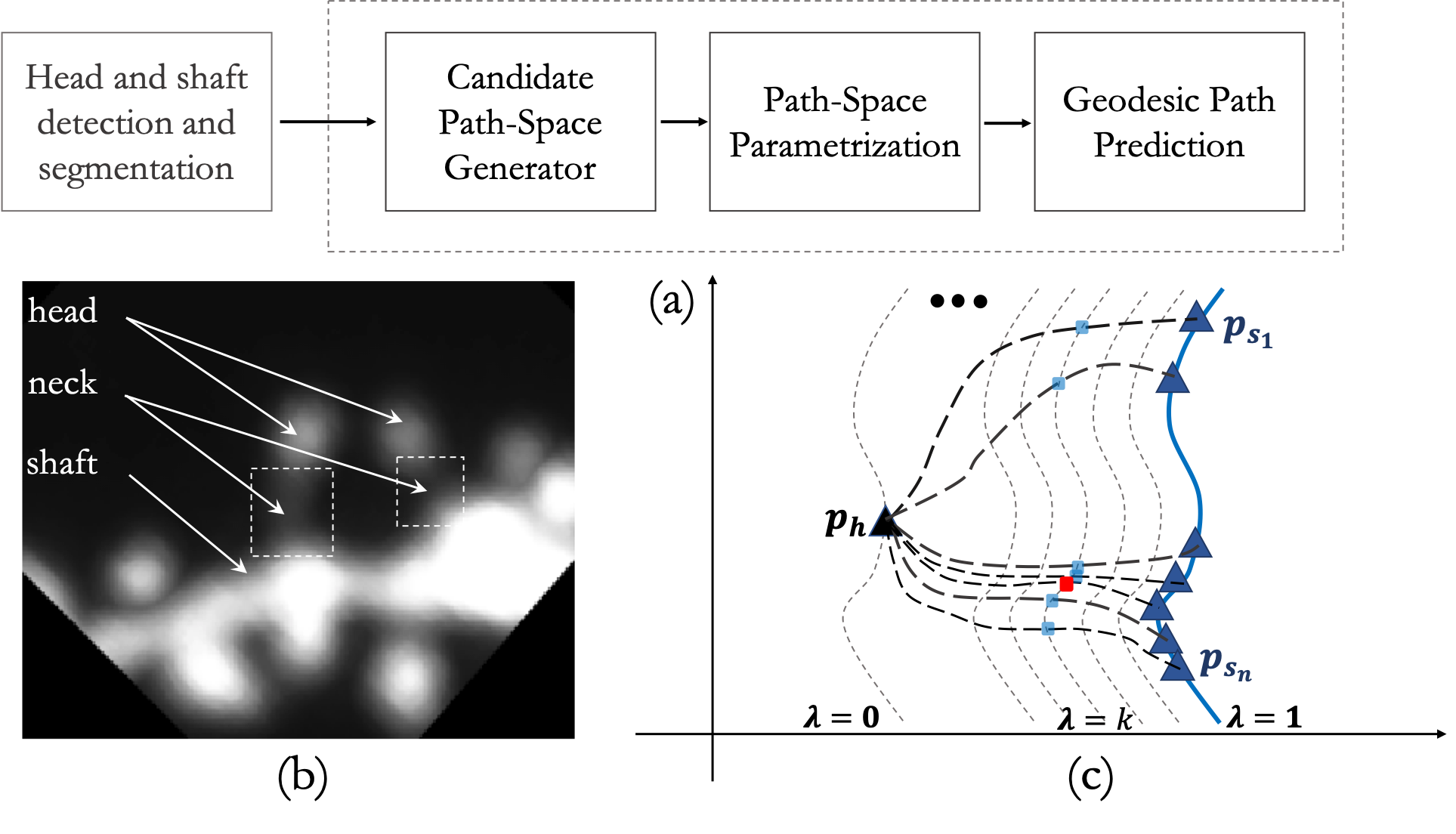} 
	\end{tabular}
	\caption[illustration]{(a) Illustration of the proposed algorithm (b) Spine morphology and (c) Illustration of path-space parameterization. The path points for $\lambda=k$ are shown, and the red point indicates the intrinsic median of the points.} 
	\label{fig:illus}
	\vspace{-16pt}
\end{figure}

Dendritic spines, which appear as  small membranous protrusions from a neuron’s dendritic shaft, are  critical for establishing excitatory synaptic contact in the neural circuitry. Morphology of dendritic spines change with learning and brain development, and studies have associated structural anomalies of spines to several neuro-developmental diseases\cite{danglot2012vezatin}. Spine anatomy is typically characterized by a post-synaptic density rich region called spine head, which is connected to the dendritic shaft via a membranous structure called spine neck (see Fig.~\ref{fig:fig1} and Fig.~\ref{fig:illus}(b) for illustration). Structural characteristics of the spine neck and head regions (such as geodesic neck length, head and neck measurements etc.) are critical  to further our understanding about signal transmission in the brain \cite{lagache2018electrodiffusion}, which makes it crucial to study their anatomical properties during brain development. 

Deciphering the relationship between the anatomy and function of neuronal components calls for large scale morphological analysis of the individual structures. Automatic segmentation of dendritic spines in fluorescent microscopy is a challenging problem especially when the spine head appears detached from the neuronal shaft due to signal attenuation in the neck membrane. Existing segmentation techniques often assume proximity of the spine head to the dendritic shaft. While this is valid for stubby spines with short necks, \textit{thin} dendritic spines with long protrusions are exceptions to this rule, and consequently result in disjoint  components via traditional segmentation methods (see Fig.~\ref{fig:fig1}). In this paper we introduce an unsupervised model to reconstruct the spine morphology by estimating the optimal path\cite{cohen1997global,chen2018minimal} which associates the detached spine head to the corresponding dendritic shaft.
 The motivation for this work, and the technical details are discussed in the following section.
%
\vspace{-6pt}
\section{Background and Motivation} 
\vspace{-6pt}
In this study we primarily analyze thin dendritic spines with poorly illuminated neck membranes which are difficult to segment automatically. An example is shown in Fig.~\ref{fig:fig1}(b), where the segmentation method\cite{l2s} fails to encapsulate the spine morphology as a single connected structure.
To address such instances of incomplete segmentation, a few techniques have been proposed to analyze the association between the disjoint components. The graph-theoretic technique Tree2Tree\cite{mukherjee2013tree2tree2} uses geometric features to determine the connectivity between the various subcomponents of an object. The fundamental challenge in adapting this solution to our problem is that spine heads are bulbous protrusions which cannot be modeled as tree shaped objects, which is a prerequisite for \cite{mukherjee2013tree2tree2}. Similarly, the variational method in \cite{tuff} can only handle structural discontinuities which are in close proximity to each other.
Another popular approach to solve this inter-structure association issue is to explore the local neighborhood of the detached head to identify connectivity clues. Su et al.\cite{su2014novel} use directional filters to detect and segment dendritic spines, although this method does not explicitly address the association problem for detached cases. Erdil et al.\cite{8759273} use deformable models with learned shape priors to segment spines from two photon microscopy images. However, due to the relative scarcity of training samples for thin spines, such supervised models would suffer from significant class imbalance. Janoos et al.\cite{janoos2009robust}   suggest iterative surface evolution to propagate the spine head until it connects with the dendritic shaft. While this method can handle detached spines, it is sensitive to false positives due to clutter as the curve propagation terminates immediately on reaching the nearest segmented object in the field of view. In summary, precise morphological reconstruction of thin dendritic spines remains a non-trivial task which demands further investigation. 
\vspace{-8pt}
\section{Method}
\vspace{-6pt}
The key contribution of this work is to estimate the entire spine morphology from an incomplete segmentation.
The broader topics of automatic spine detection, or neurite segmentation\cite{tuff,meijering2010neuron} are not addressed here. We also assume the availability of a reliable object detection technique (such as  \cite{xiao2018automated}) to identify the spine head and the corresponding dendritic shaft. 
The initial localizations of the spine head and the dendritic shaft are used to segment the structures using the active contour model Legendre Level Set (L2S)\cite{l2s} . While L2S can reliably reconstruct both stubby and mushroom spines, the elongated, thin varieties pose challenge due to significant signal attenuation at the neck membrane. Our objective is to identify the optimal curve to represent the undetected spine neck.
A schematic representation of the algorithm is shown in Fig.~\ref{fig:illus}(a), and our technique is formally presented next.
\vspace{-10pt}
\subsection{Problem formulation and mathematical details}
\vspace{-4pt} 
We denote the centroid of the segmented spine head by the coordinate $p_h = \left(x_h, y_h\right) \in \Omega$, where $\Omega\subset\mathbb{R}^2$ defines the domain of the (2D) image $g:\Omega\mapsto\mathbb{R}$. The spine neck centerline joining the head and the shaft is represented by a Lipschitz regular parametric curve $C:\left[0,1\right]\mapsto\Omega$. Finding the right parameterization method is essential to our solution, and this will be detailed in the following subsections. 
\vspace{-8pt}
\subsubsection{Path-Space and Minimal Paths}
\vspace{-4pt} 
A \textit{path-space}\cite{basu2011path2path} of a collection of curves between the spine head $p_h$ and any point $p\in \Omega$ on the dendritic shaft boundary is defined as follows:
\begin{align}
\mathcal{C}_{p_h,p}=\{C_j| C_j\left(0\right)=p_h \, \text{and}\, C_j\left(1\right)=p\} \, \forall j\in\mathbb{N}
\label{eq:path_space}
\end{align}
We seek the optimal path $\hat{f}\in \mathcal{C}_{p_h,p}$ from this path-space by optimizing a suitable objective which penalizes imprecise reconstructions.
When both the source ($p_h$) and the terminal ($p$) points of the curve are known, Cohen and Kimmel\cite{cohen1997global} describe a fast algorithm to compute a smooth curve as a solution to the following initial value problem:
\begin{align}
||\nabla \mathcal{U}_h|| &= \mathcal{P}(x,y) \quad \text{with} \quad \mathcal{U}_h(p_h) = 0 \label{eq:fmm}\\ 
\text{where} \,\,\, \mathcal{U}_h(p) &= \inf_{\mathcal{C}_{p_hp}} \int_{0}^{1}\underbrace{\Big(w+\mathcal{P}\big(C\left(\lambda\right)\big)\Big)}_{\tilde{\mathcal{P}}}|C'(\lambda)|d\lambda \nonumber
\vspace{-18pt}
\end{align}
Here $C\in\mathcal{C}_{p_hp}$, and the geodesic distance map $\mathcal{U}_h:\Omega\mapsto\mathbb{R}$ defines the level set function of the arrival time of a wavefront propagating from $p_h$ according to $C_t = \mathcal{\tilde{P}}^{-1}\mathcal{N}$. Here $C_t$ is the partial derivative of the curve with respect to the pseudo time $t$ and $\mathcal{N}$ is the curve unit normal vector. 
$\tilde{\mathcal{P}}=w+\mathcal{P}$ is the inverse speed function for curve propagation. The positive scalar $w$ promotes smoothness, and $\mathcal{P}$ is typically designed such that the valleys of the function would correspond to the minimal geodesic. One such realization is  $\mathcal{P}(x,y)=e^{-\mu g(x,y)}$. We choose $w=0.01$ as prescribed in\cite{mirebeau2019hamiltonian}, and the rational for selecting $\mu$ will be explained later.  
\vspace{-8pt}
\subsubsection{Generative model for   candidate path-space}
\vspace{-4pt} 
By performing gradient descent on the potential function $\mathcal{U}_h$, it is theoretically feasible to extract all possible geodesic paths in  $\mathcal{C}_{p_h,p}$, although such a brute force approach is computationally restrictive. To circumvent this issue, we propose the following algorithm. Each point on the dendritic shaft boundary set $S=\{p_s\}$ is associated with an arrival time $t_s$ of the front propagated from $p_h$ by solving eq.~\ref{eq:fmm} via fast marching\cite{mirebeau2019hamiltonian}. A simple path prediction strategy which is similar to\cite{janoos2009robust} is to select the curve terminal $p^*_s \in S$ such that $t^*_s = \min \{t_s\}$, but this is quite sensitive to 
clutter and the parameter $\mu$, and may result in suboptimal solutions(see Fig.~\ref{fig:compare}).
To avoid such problems, we propose the following strategy.
For each point $p_s\in S$, we associate a weight function $w_s\in \left[0,1\right]$ as follows:
\begin{align}
w_s = 1-\left(t_s - t_-\right)/\left(t_+ - t_-\right)
\label{eq:sampling}
\end{align}
Here  $t_+$ and $t_{-}$ are the maximum and minimum values in $T_s=\{t_s\}$. We implement a stochastic sampling strategy to compute a set  $S_c\subset S$ of most probable curve end points. First, a random subset of points are selected from $S$. Then each sample $p_s\in S$ is added to the candidate set $S_c$ with probability $w_s$. This  sampling (without replacement) procedure is continued until a specified number ($n$) of path terminals are accumulated. The candidate path terminals are used to generate the candidate subspace $\mathcal{C}_n=\{C_1,\ldots C_n\}\subset\mathcal{C}_{p_h,p}$ where $n=|S_c|$. This candidate path-space serves as the observed data point set for the path estimation problem.
\vspace{-8pt}
\subsubsection{Path-space parameterization}
\vspace{-4pt} 
The geodesic path prediction is  expressed as $\hat{f} = \chi\left(\mathcal{C}_n\right)$, where $\chi$ is a predictor functional which needs to be estimated from the observation $\mathcal{C}_n$ based on a specific optimality criteria.
However, in order to define a set of algebraic operations over $\mathcal{C}_{p_h,p}$, it is first necessary to perform a suitable re-parameterization of the path-space. This is done by defining the curve parameter $\lambda\in \left[0,1\right]$ for each point $(x,y)$ on a curve  $C \in \mathcal{C}_{p_h,p}$ as follows:
\begin{align}
\lambda = 1 - \phi(x,y)/\phi(x_h, y_h) 
\label{eq:parameter}
\end{align}
Here $\phi$ denotes the level set function of the segmented dendritic shaft and is defined to be $\phi(x,y)\geq 0$ $\forall (x,y)$ outside the shaft segmentation\cite{l2s}, and the shaft boundary represents the zero level set of the function. From eq.~\ref{eq:parameter}, we can infer that $\lambda=1$ when $p\in S$ and $\lambda=0$ if $p=p_h$, which normalizes the path-space such that $C(0)=p_h$ and $C(1)=p_s$, $\forall C\in\mathcal{C}_{p_h,p}$ and $p_s\in S$ (see Fig.~\ref{fig:illus}(c) for illustration). This parameterization strategy enables standard mathematical operations on the elements of the path-space. Although eq.~\ref{eq:parameter} implicitly assumes simple curves without self loops, this is indeed an appropriate assumption since in practice the geodesic path samples obtained via eq.~\ref{eq:fmm} are typically regularized to be smooth and self-intersecting geodesic paths are rarely encountered.
\vspace{-8pt}
\subsubsection{Geodesic path prediction}
\vspace{-4pt}
We introduce a probabilistic framework to estimate the optimal curve to represent the spine neck centerline. In this paradigm, the curve parameter $\lambda$ is considered to be a realization of some random variable $\Lambda$. Similarly, the curve coordinates are assumed to be the realizations of a stochastic response variable $Y\in\mathbb{R}^2$. Both $\Lambda$ and $Y$ are assumed to be associated with a differentiable probability density function. We wish to obtain the optimal functional $\hat{f}$ by minimizing the expected value of the stochastic function defined as follows:
\begin{align}
\hat{f} &= \underset{f}{\argmin\,} \mathbb{E}_{Y\Lambda}\big[||Y-f(\Lambda)||_1\big] \nonumber \\
\text{or,}\,\,\hat{f}(\lambda)	 & = \underset{f}{\argmin\,} \mathbb{E}_{\Lambda}\mathbb{E}_{Y|\Lambda}\big[||Y-f(\Lambda)||_1\big|\Lambda=\lambda\big] 
\label{eq:min1}
\end{align}
Here $\mathbb{E}_\Lambda\left[.\right]$ computes the expected value of a random variable $\Lambda$. The joint expectation of $Y$ and $\Lambda$ is represented by $\mathbb{E}_{Y\Lambda}\left[.\right]$, and $\mathbb{E}_{Y|\Lambda}\left[.\right]$ is the conditional expectation operator. The solution to eq.~\ref{eq:min1} is obtained by minimizing the function in a point-wise fashion\cite{elsl}. Formally,
\begin{align}
\hat{f}(\lambda) &= \underset{c\in \mathbb{R}^2}{\argmin\,} \mathbb{E}_{Y|\Lambda}\big[||Y-c||_1\big|\Lambda=\lambda\big]
\label{eq:min2}
\end{align}
The spine neck is reconstructed by evaluating the path coordinates at $N$ sample points $\lambda_i$ of the continuous curve parameter space. Formally, the discrete equivalent of the optimal path is evaluated for each $\lambda_i$ as
\begin{align}
\hat{f}(\lambda_i)\approx\chi\Big(C_1(\lambda_i),\ldots,C_n(\lambda_i)\Big), i=1,\ldots,N
\label{eq:chi}
\end{align}
The function $\chi$ estimates the point-wise minimizer of eq.~\ref{eq:min2} by computing the intrinsic median\cite{elsl} of the sample path points evaluated at $\lambda=\lambda_i$ (see Fig.~\ref{fig:illus}(c)). Finally, a smooth representation of the spine neck path is obtained by regressing a cubic spline to the discrete set of path coordinates.
\vspace{-10pt}
\subsection{Discussion}
\vspace{-4pt}
Eq.~\ref{eq:min1} penalizes the  $\ell_1$ norm  of the expected reconstruction error, and the minimizer is realized in eq.~\ref{eq:chi} by computing the point-wise intrinsic median of the observed candidate path coordinates. Finding the intrinsic median requires an ordering of the path points on the level sets of $\phi$, which is trivial to compute since the path-space is standardized by the aforementioned parameterization procedure. The median order statistic makes this solution robust to outliers in path prediction, and the false positives are further reduced due to stochastic sampling of the path terminals. 
\vspace{-6pt}
\section{Experiments}
\begin{figure}[t]
	\setlength{\tabcolsep}{1pt}
	\centering
	\begin{tabular}{cccc}
		\includegraphics[width=.24\linewidth]{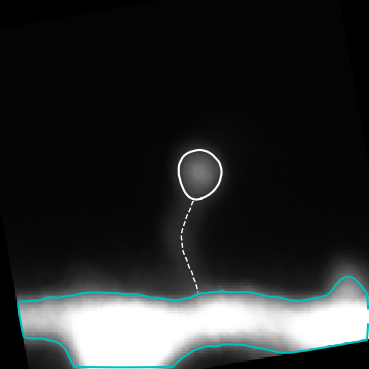} &
		\includegraphics[width=.24\linewidth]{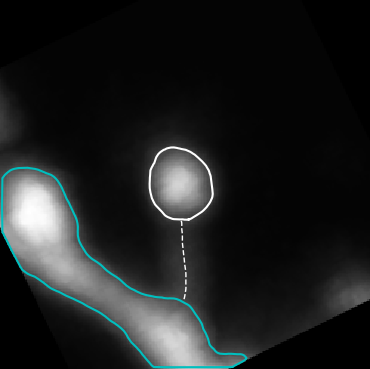} &		
		\includegraphics[width=.24\linewidth]{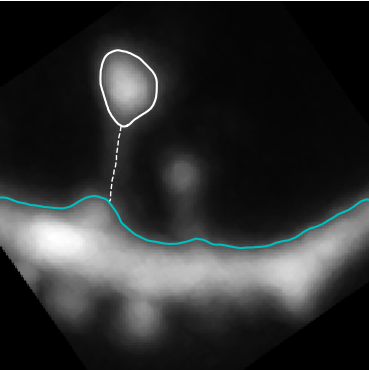} &
		\includegraphics[width=.24\linewidth]{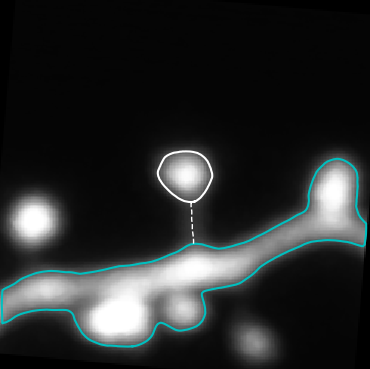} \\
		\includegraphics[width=.24\linewidth]{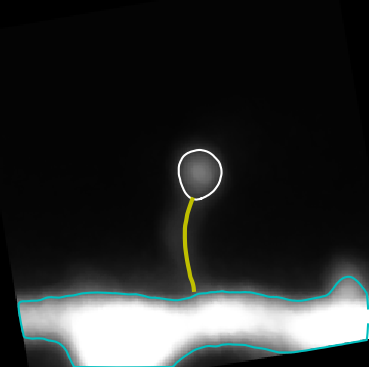} &
		\includegraphics[width=.24\linewidth]{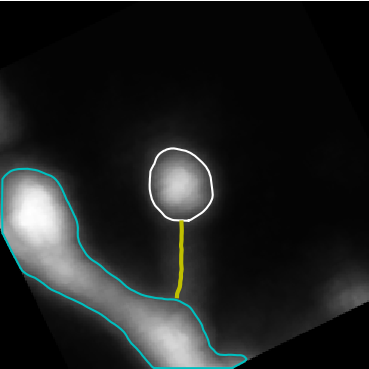} &		
		\includegraphics[width=.24\linewidth]{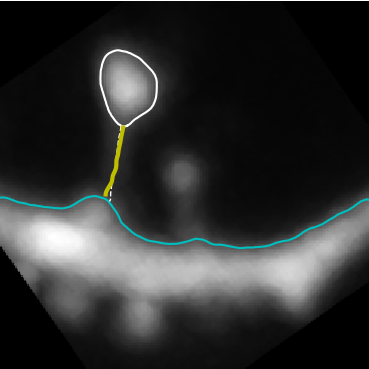} &
		\includegraphics[width=.24\linewidth]{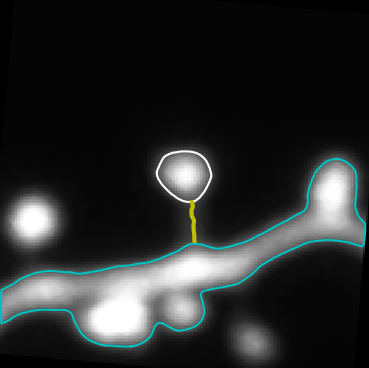} 
	\end{tabular}
	\caption[illustration]{A few cases of path prediction for thin spines are shown here. The white dotted curve indicates ground truth, and the predicted path is shown in yellow. The cyan contour shows the dendritic shaft segmentation, which is detached from the shaft head (white contour)} 
	\label{fig:display}
	\vspace{-16pt}
\end{figure}

\begin{figure}[h]
	\setlength{\tabcolsep}{16pt}
	\centering
	\begin{tabular}{cc}
		\includegraphics[width=.35\linewidth]{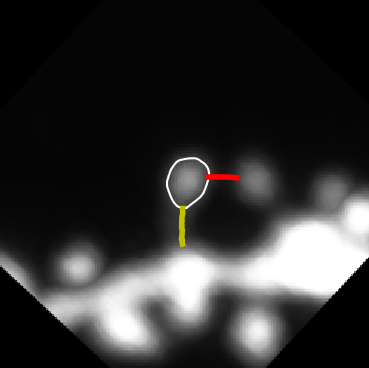} &
		\includegraphics[width=.348\linewidth]{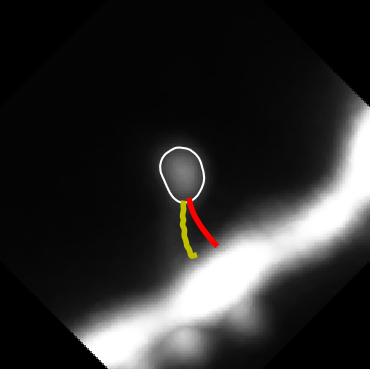} \\
	\end{tabular}
	\caption[illustration]{The path prediction strategy which selects the closest shaft-point\cite{janoos2009robust} is shown in red, and the results due to the  proposed method is shown in yellow. It may be inferred from the results that the proposed technique is more robust to outliers.} 
	\label{fig:compare}
	\vspace{-16pt}
\end{figure}

\vspace{-6pt}
We demonstrate our algorithm on the dataset of two photon microscopy images of dendritic spines presented in\cite{ghani2017dendritic}. The results are evaluated primarily on thin spines where the segmentation tool L2S was unable to segment and associate the head and the dendritic shaft. 
A few illustrative examples are shown in Fig.~\ref{fig:display} for qualitative assessment. The images in Fig.~\ref{fig:display} are characterized by significant signal attenuation in the  neck region, and consequently the head and the shaft are rendered as separated objects. The second row of Fig.~\ref{fig:display} shows the spine neck prediction results using our method.
An important contribution of this work is to introduce stability and robustness in path prediction. Fig.~\ref{fig:compare} demonstrates two representative examples where the proposed solution (shown in yellow trace) is able to extract the precise morphology of the spine neck, as opposed to the front propagation method in \cite{janoos2009robust} (in red) which is susceptible to sub-optimal local solutions.

To quantitatively analyze the performance of our solution, we measure the accuracy of tracing the spine neck centerline using the mean absolute error metric\cite{tuff} between a  discrete reconstruction $U=\{u_1,\ldots,u_m\}$ and the corresponding manual annotation $V=\{v_1,\ldots,v_n\}$ $(u_i, v_j \in \mathbb{R}^2)$. This is defined as $MAE = \dfrac{1}{m}\sum_{i=1}^m \min_j |u_i - v_j| +  \dfrac{1}{n}\sum_{i=1}^n \min_j |v_i - u_j|$. We measure the quantitative results against the strategy in \cite{janoos2009robust} which connects the spine head to the nearest shaft point. Experiments were performed on a set of 50 images, and the average MAE due to our method is measured to be $4.2\pm3.2$ pixels, which is an improvement of $12.5\%$  over the competitor which yields an average MAE of $4.8\pm5.4$ pixels. Furthermore, using our methodology we observe $40\%$ lower standard deviation of reconstruction error, which testifies to the robustness of our methodology.
We also analyzed the sensitivity of our algorithm to the choice of the parameter $\mu$ which defines the curve propagation speed. Experimental analysis reveal that the proposed method is significantly stable for a wider range of $\mu\in\left[3.7, 10\right]$, whereby the technique in \cite{janoos2009robust} exhibits oscillatory behavior unless the $\mu$ is precisely selected to be in the range $\mu\in\left[8.7,10\right]$. This emphasizes the robustness of our methodology to  model parameter selection.

\vspace{-8pt}
\section{Conclusion}
\vspace{-6pt}
To summarize, we have introduced a formal methodology for accurate morphological reconstruction of thin dendritic spines. The optimization criteria for path estimation ensures robustness of our solution, since the point-wise estimate via intrinsic median (eq.~\ref{eq:chi}) efficiently rejects outliers. Additionally, due to the intelligent path sampling strategy, we reduce bias due to local minima issue. The experimental results have been demonstrated on two-dimensional images, but the algorithm can be extended to three dimensional analysis with little modifications in the processing pipeline. This will be explored in more detail in our future studies.

\bibliographystyle{IEEEbib}
{	\footnotesize
	\bibliography{refs}}

\end{document}